\documentclass[12pt]{aastex}
\usepackage{emulateapj5}
\usepackage{epsfig}
\usepackage{graphicx}

\def\ltsima{$\; \buildrel < \over \sim \;$}
\def\lsim{\lower.5ex\hbox{\ltsima}}
\def\gtsima{$\; \buildrel > \over \sim \;$}
\def\gsim{\lower.5ex\hbox{\gtsima}}

\shorttitle{High Redshift Supernova Rates}
\shortauthors{Dahlen et al.}

\begin{document}


\title{High Redshift Supernova Rates}


\author{
Tomas Dahlen\altaffilmark{1},
Louis-Gregory Strolger\altaffilmark{1},
Adam G. Riess\altaffilmark{1},
Bahram Mobasher\altaffilmark{1,2},
Ranga-Ram Chary\altaffilmark{3},
Christopher J. Conselice\altaffilmark{4},
Henry C. Ferguson\altaffilmark{1},
Andrew S. Fruchter\altaffilmark{1},
Mauro Giavalisco\altaffilmark{1},
Mario Livio \altaffilmark{1},
Piero Madau\altaffilmark{5},
Nino Panagia\altaffilmark{1,2},
and John L. Tonry\altaffilmark{6}
}



\altaffiltext{1}{Space Telescope Science Institute, 3700 San Martin Dr., Baltimore, MD 21218; dahlen@stsci.edu; strolger@stsci.edu; ariess@stsci.edu; mobasher@stsci.edu; ferguson@stsci.edu; fruchter@stsci.edu; mauro@stsci.edu; mlivio@stsci.edu; panagia@stsci.edu}
\altaffiltext{2}{Affiliated with the Space Telescope Division of the European Space Agency, ESTEC, Noordwijk, The Netherlands}
\altaffiltext{3}{Spitzer Science Center, MS220-6, California Institute of Technology, Pasadena, CA 91125; rchary@caltech.edu}
\altaffiltext{4}{California Institute of Technology, Pasadena, CA 91125; cc@astro.caltech.edu}
\altaffiltext{5}{Department of Astronomy and Astrophysics, University of California, Santa Cruz, 1156 High Street, Santa Cruz, CA 95064; pmadau@ucolick.org}
\altaffiltext{6}{Institute for Astronomy, University of Hawaii, 2680 Woodlawn Drive, Honolulu, HI 96822; jt@ifa.hawaii.edu}


\begin{abstract}
We use a sample of 42 supernovae detected with the Advanced Camera for Surveys on-board the Hubble Space Telescope as part of the Great Observatories Origins Deep Survey to measure the rate of core collapse supernovae to $z\sim$~0.7 and type Ia supernovae to $z\sim$~1.6. This significantly increases the redshift range where supernova rates have been estimated from observations.

The rate of core collapse supernovae can be used as an independent probe of the cosmic star formation rate. Based on the observations of 17 core collapse supernovae, we measure an increase in the core collapse supernova rate by a factor of $\sim$~1.6 in the range $0.3<z<0.7$, and an overall increase by a factor of $\sim$~7 to $z\sim$~0.7 in comparison to the local core collapse supernova rate. The increase in the rate in this redshift range in consistent with recent measurements of the star formation rate derived from UV-luminosity densities and IR datasets.

Based on 25 type Ia supernovae, we find a SN Ia rate that is a factor $3-5$ higher at $z\sim$~1 compared to earlier estimates at lower redshifts ($z<0.5$), implying that the type Ia supernova rate traces a higher star formation rate at redshifts $z>1$ compared to low redshift. At higher redshift ($z\gsim$1), we find a suggested decrease in the type Ia rate with redshift. This evolution of the Ia rate with redshift is consistent with a type Ia progenitor model where there is a substantial delay between the formation of the progenitor star and the explosion of the supernova. Assuming that the type Ia progenitor stars have initial main sequence masses $3M_{\odot}<M<8M_{\odot}$, we find that $5-7$\% of the available progenitors explode as type Ia supernovae.
\end{abstract}




\section{Introduction}
The rates of supernovae (SNe) at different redshifts provide important information about the evolution of a number of physical processes over cosmic times. As core collapse SNe (i.e., type II and Ibc supernovae, hereafter CC SNe) originate from massive short-lived stars, the rate of these events should reflect on-going star formation and therefore offers an independent way to determine the cosmic star formation rate (SFR). Furthermore, the SN rate (SNR) directly probes the metal production at different cosmological epochs.  

Type Ia SNe do not directly follow the SFR since there is a delay between the formation of the progenitor star and the explosion of the SN. Constraining this delay time is important for a better understanding of the processes leading to the type Ia explosions, and therefore essential for the usefulness of type Ia SNe as cosmological distance indicators. With accurate measurements of the SN Ia rate, it should be possible to set constraints on this poorly known delay time as discussed in e.g., Madau, Della Valle, \& Panagia (1998) and Dahlen \& Fransson (1999). A first attempt to constrain the delay time using observations is presented in Gal-Yam \& Maoz (2004) and Maoz \& Gal-Yam (2004). 

Using SNRs as evolutionary probes requires the rates to be measured over cosmological distances where the properties of the Universe, e.g., the SFR($z$), are believed to change by a significant fraction. In practice, this means to redshifts $z\sim$1, which is higher than has previously been feasible. Local and nearby SN Ia rates have been measured in the field by numerous surveys, e.g., at redshifts $<z>\sim0.01$~(Cappellaro, Evans, \& Turatto 1999), $<z>=0.1$~(Hardin et al. 2000), $<z>=0.11$~(Strolger 2003), and $<z>=0.114$~(Reiss 2000). Similarly, at larger distances, rates have been measured at $<z>=0.38$~(Pain et al. 1996), $<z>=0.46$~(Tonry et al. 2003), and $<z>=0.55$~(Pain et al. 2002). While rates at $z\gsim 0.4$~are in general higher than the local rates, these measurements are still consistent with a constant type Ia rate out to $z\sim$0.6 (Tonry et al. 2003). In clusters of galaxies, rates have been estimated at $<z>=0.25$~and $<z>=0.9$~(Gal-Yam, Maoz, \& Sharon 2002).

By contrast, CC SNe are typically $\sim2$~mag fainter and searches are affected by severe selection biases, thus only locally determined rates of CC SNe exist (e.g., Cappellaro et al. 1999; Strolger 2003).   

The Great Observatories Origins Deep Survey (GOODS, Giavalisco et al. 2004a) has offered an unprecedented  opportunity to obtain a sample of high redshift SNe, ideal for studying the evolution of the SNR with redshift. Using the Advanced Camera for Surveys (ACS) on-board the Hubble Space Telescope (HST), GOODS has detected 42 SNe in the redshift range $0.2\lsim z\lsim 1.6$. Repeated observations of two fields, combined with the high spatial resolution of the ACS, have resulted in a sample with well understood detection efficiency and systematics. In this paper, we derive and discuss the rates of CC and type Ia SNe based on this unique sample. In an accompanying paper (Strolger et al. 2004), we describe details of the search, including data reduction procedures, SN finding and classification, selection biases, detection efficiency and completeness. This paper also presents a detailed investigation of SN Ia progenitor delay time distributions derived from the GOODS SN sample. The SN sample, including magnitudes and redshifts, is presented in Strolger et al. and Riess et al. (2004). Constraints on cosmological parameters using the GOODS SN sample are presented in Riess et al.

The data are presented in \S~2. In \S~3 we describe how the SN rates are determined from observations. Results are given in \S~4, followed by an investigation of the relation between star formation and SN rates in \S~5. Results are discussed in \S~6. Finally, conclusions and a summary are given in \S~7.  

Throughout this paper, we adopt a flat lambda dominated cosmology ($\Omega_{\Lambda}=0.7, \Omega_M=0.3$) and a Hubble constant $H_0=70$~km s$^{-1}$~Mpc$^{-1}$. Magnitudes are given in the Vega based system.

\section{Data}
\subsection{Observations}
During the GOODS campaign, two fields (HDF-N and CDF-S) were observed using HST/ACS during five epochs separated by $\sim$45 days. Both fields were observed with multiple pointings, covering an effective area $\sim$150 sq. arcmin per field. The search was conducted using observations in the $z$-band (F850LP filter), with additional observations in $V$ and $i$ (F606W, F775W), allowing colors to be calculated. For a number of high redshift SN Ia candidates, near IR photometric (HST/NICMOS) and spectroscopic (VLT, Keck) follow up was conducted, as reported in Riess et al. (2004) and Strolger et al. (2004)

When searching for SNe, we subtracted two consecutive epochs creating a residual image, which thereafter was both automatically and visually scanned for objects. Each area was independently investigated by two teams. Completeness was estimated by adding artificial point sources with different magnitudes to the real images. The resulting completeness limit is $m_{F850LP} = 25.9$, defined as the magnitude where 50\% of these sources where recovered. 

\subsection{The Supernova Sample}
The GOODS SN sample consists of 42 SNe detected in both CDF-S and HDF-N. For 29 of these, we have spectroscopic redshift determinations from either the SNe themselves, or the host galaxies, while we have photometric redshifts (Mobasher et al. 2004) for an additional 12 SNe. We only lack redshift for one SN with no apparent host galaxy.

We wish to differentiate between type Ia and CC SNe when calculating rates since these types originate from different progenitors and explosion mechanisms. It is therefore essential to conduct accurate type determination for the full sample. We divide the SNe into types using all available information including spectra, redshifts of SNe or host galaxies, magnitudes, colors and light curve shapes. We find that 25 SNe are consistent with being type Ia, with the remaining 17 being CC SNe. We are most confident in the type determination of 20 of the Ia and 7 of the CC SNe. For the remaining ones, the type determination is less certain and a few may be misclassified. When calculating rates, we account for this by including possible misclassifications in the systematic errors. 

\section{Determining SN Rates from Observations}
In this paper, we express SNRs in units of number of exploding SNe per rest-frame year and comoving volume element (i.e., yr$^{-1}$Mpc$^{-3}$). Alternatively, rates can be expressed in supernova units, SNu's (1 SNu = 1 SN per 100 year per 10$^{10}L^B_{\odot}$), which are normalized to the $B$-band luminosity density, and are preferentially used for local measurements. We do not use SNu's because this introduces further uncertainty in surveys covering cosmological distances where the rest-frame $B$-band luminosity and its evolution has to be determined to provide proper rates in SNu's. When we compare local rates given in SNu's with rates per volume, we convert units assuming a local $B$-band luminosity density 2.0$\times 10^8hL_{\odot}$Mpc$^{-3}$, approximately evolving as $(1+z)^{1.9}$ at $z<$1 (Strolger 2003). The redshift correction is, however, small since the highest redshift we convert is $z\sim 0.1$. 

We use Monte Carlo (MC) simulations to calculate the underlying SNR that is consistent with observations. This is done separately for CC and type Ia SNe. We start by assuming an input SNR and use this to calculate the number of SNe exploding each year within a volume given by the field-of-view and the redshift range of interest. Each SN is assigned a peak magnitude, a random epoch on the light curve, a host galaxy extinction and a redshift. The apparent magnitude is thereafter calculated at this epoch on the light curve, as well as at an earlier epoch separated in rest-frame by $45/(1+z)$ days, corresponding to the 45 day spacing between observations. By subtracting the fluxes between the two epochs, we derive a detection magnitude, $\Delta m$, and the direction of the magnitude change, i.e., if the SN is rising or declining. This gives us a list of simulated SNe with known detection magnitudes and redshifts. The MC simulation is repeated 10,000 times to calculate the mean number of detectable SNe for a specified search set-up.  Finally, to determine the observed rate, we adjust the input rate to match the number of detected SNe in a chosen redshift bin.

Calculating the apparent magnitude of the SNe in the simulations includes specifying a number of SN characteristics. After the following general formula, we specify the CC and SN Ia characteristics in the two next subsections. The apparent magnitude in the observed filter $F850LP$ for a SN at redshift $z$ which is at time $t$ relative to the light curve peak in the observers frame is given by
\begin{center}
\begin{math}
m_{F850LP}(t,z)=M_{peak,\lambda} + \Delta M_{\lambda}[t\times(1+z)^{-1}]
\end{math}
\end{center}
\begin{equation}
+~DM(z)-K^{\lambda}_{F850LP}[z,t\times(1+z)^{-1}]+{\rm A}_{\lambda}
\end{equation}
where $M_{peak,\lambda}$ is the peak absolute magnitude in a rest-frame filter $\lambda$, $\Delta M_{\lambda}$ is the light curve decline relative to the peak, $DM(z)$ is the distance modulus. The K-correction, $K^{\lambda}_{F850LP}$, is calculated using the formalism in Kim, Goobar, \& Perlmutter (1996) and includes information on the spectral SED of the SN together with appropriate filter transmission curves and QE for the detector used. Finally, A$_{\lambda}$ is the extinction in the SN host galaxy. This is further described below. 
\subsection{Core Collapse SN Characteristics}
CC SNe can be divided into a number of different subtypes, characterized by
e.g., light curve decline, spectral features or peak magnitude. We adopt the division into subtypes given in Richardson et al. (2002). Here CC SNe are divided into Ibc, IIL, IIP and IIn, where the Ibc and IIL types are further divided into a normal and a bright population, while the IIP type is divided into a normal and a faint population. A summary of peak $B$-band magnitudes and corresponding dispersions are given in Table 1. We use the intrinsic fractions of subtypes given in Dahlen \& Fransson (1999), with the addition of 1\% of the SNe assigned to a bright population as proposed by Richardson et al. This bright population is further divided so that half are Ibc and half IIL. Note that the division into different types is highly uncertain. Also, the intrinsic fractions may change with redshift (metallicity). In \S~5.1.1, we investigate how the subdivision into subtypes affects the results.

To obtain K-corrections, we use the approach in Dahlen \& Fransson (1999), where the SN spectra are characterized by black bodies with temperatures evolving from $\sim$25,000K near explosion to $\sim$5,000K at late stages. We use these models since there does not exist a full sample of observed CC SN spectra of different subtypes and epochs on the light curve covering a sufficiently large range in wavelength to calculate K-corrections. For particular SNe (e.g., 1999em) with good observational coverage, we find that the blackbody approximation and observed spectra yield similar K-corrections, with deviations mostly $<$ 0.1 mag. This is typically less than the photometric uncertainty in the observed CC SNe, justifying the use of the blackbody approximation. Light curves for Ibc, IIL, and IIP are taken from Filippenko (1997), while for IIn, we use a light curve intermediate between IIL and IIP.

CC SNe should preferentially occur in star bursting regions within the host galaxy. We therefore use a starburst extinction law (Calzetti et al. 2000) when calculating absorption of the SN due to dust in the host galaxy. Furthermore, we use a mean color excess $E(B-V)$=0.15. The details of dust extinction, especially in star forming regions, are, however, highly uncertain and in \S~5.1.1, we thoroughly investigate how different assumptions about the extinction affects results. 

Each SN is placed in a disk with random inclination. The extinction is thereafter calculated by integrating the light path through the host galaxy using the recipe in Hatano, Branch, \& Deaton (1998). This leads to a distribution of extinctions that peaks at low values, A$_{\rm B}~<~0.1$, but that also shows a tail with high extinction values (A$_{\rm B}~>~5-10$), originating from highly inclined host galaxies.

\subsection{Type Ia Characteristics}
To characterize type Ia SNe, we use the peak magnitude distributions in Tonry et al. (2003). Here type Ia SNe are divided into three Gaussians with peak magnitudes $M_B$=--19.6, --19.3, --17.8 and dispersions $\sigma$=0.30, 0.45, and 0.50, respectively (see Table 1). The SNe are distributed within these three distributions so that 20\% are 'bright', 64\% are 'normal', and the remaining 16\% are 'dim' (Li et al. 2001).

K-corrections are calculated using a set of observed spectral templates from SN1994D, observed at days --5, 0, 5, 9, 17, and 55 relative to peak in light curve. We interpolate between these spectra to get K-corrections at intermediate epochs on the light curve. We use a mean $B$-band light curve taken from Riess et al. (1999).

Absorption in the host galaxy is calculated as in Hatano et al. (1998). We divide the SN Ia so that 25\% are placed in a bulge, while the remaining are placed in a disk as suggested by Tonry et al. (2003). This is consistent with the fraction $\sim$30\% found by Farrah et al. (2002), as well as preliminary results from the GOODS SN sample. The scale height used for the disk in Hatano et al. (1998) is larger than for the CC SNe, leading to lower mean absorption for SN Ia. For both bulge and disk components, the distribution of host galaxy $B$-band absorption has a very dominating peak at low extinctions (A$_{\rm B}~<~0.1$). There is, however, also a narrow tail out to high extinctions (A$_{\rm B}~\gsim~4$), especially for SNe occurring in inclined disks. The overall distribution of extinctions can be approximated by an exponential, $P$(A$_{\rm B})\propto e^{-{\rm A_B}}$, consistent with the observed distribution of SN Ia absorption in Jha et al. (1999).

\section{Results}
\subsection{Observed Core Collapse Rates}
Using the method described above, we calculate the CC SNR in two redshift bins $0.1\le z<0.5$~and $0.5\le z<0.9$, which is the first measurement of CC SN rates at cosmological distances. The resulting rates are 2.51$~^{+0.88~+0.75}_{-0.75~-1.86}$~and 3.96$~^{+1.03~+1.92}_{-1.06~-2.60}$ in the two bins, respectively (Table 2). Rates are given in units of $10^{-4}$yr$^{-1}$Mpc$^{-3}h_{70}^3$. First errors quoted represent 68.3\% confidence intervals, and are calculated from a distribution of rates based on 10,000 MC simulations. The 95\% confidence interval is well represented by two times the quoted statistical errors. The second quoted errors are systematic and include e.g., the possibility of misclassification of SN type. Systematic errors are further discussed in \S 5.1.1. 

The measured rates are higher by factors $\sim$~4 and $\sim$~7 at $<z>=0.3$~and $<z>=0.7$~compared to the local ($z\simeq 0.0$) rate $\sim 0.59$ in Cappellaro et al. (1999) (we convert the local rate from supernova units (SNu's) as described in \S~3). The rate in Cappellaro et al. is calculated using an empirical correction for inclination effects biasing against finding SNe with high extinction. Cappellaro et al. compare this method with the corrections in Hatano et al. (i.e., similar to what we use) and find results that are consistent within $\sim$15\%. The main result, that our rates at $0.3<z<0.7$~are significantly higher than the local rate, therefore seems robust, as further discussed in \S 5.1.

With the CC SNe being directly related to the SFR, the increase in the SNR reflects an increase in star formation over this redshift range, as discussed below. 

\subsection{Observed Type Ia rates}
We calculate the type Ia SNR in four redshift bins $0.2\le z<0.6$, $0.6\le z<1.0$, $1.0\le z<1.4$, and $1.4\le z<1.8$, using the MC simulations described above. Results are given in Table 2. We find a rate 0.69~$^{+0.34~+1.54}_{-0.27~-0.25}$~at $<z>=0.40$ (rates given in units $10^{-4}$yr$^{-1}$Mpc$^{-3}h_{70}^3$). This rate is somewhat higher than rates previously measured at $z\sim$ 0.4--0.5 (Pain et al. 1996, 2002; Tonry et al. 2003). The differences are within the error-bars, however, it may be possible that the derived ground-based rates are underestimated due to systematic effects, as further discussed in \S~5.2. At $<z>=0.80$, we derive a rate 1.57~$^{+0.44~+0.75}_{-0.25~-0.53}$, significantly higher than previous measurements at lower redshift. At even higher redshifts, we find a rate that is consistent with declining at $z>1$. The rate at $<z>=1.20$~is, however, still higher than local measurements. In numbers, we find rates 1.15~$^{+0.47~+0.32}_{-0.26~-0.44}$~and 0.44~$^{+0.32~+0.14}_{-0.25~-0.11}$~at $<z>=1.20$~and $<z>=1.60$, respectively. Errors represent statistical and systematic errors, as described in \S 4.1. 

While earlier measurements were still consistent with a type Ia SNR constant with redshift, our new results suggest that the SNR increases by a factor $\sim$~5 from the local Universe to a cosmic time corresponding to $z\sim$1, i.e., when the Universe was less than half its present age, and thereafter the rate shows a slight decrease toward higher redshift. Fitting a constant type Ia rate to the combined data results in a reduced chi-square $\chi^2/\nu$=2.9, equivalent to rejecting a constant rate with $\sim$99.9\% probability. The derived chi-square is based on statistical errors only, taking systematic errors into account should lower the significance.  

The increased rate to redshift $z\sim$1~is expected, since it reflects the observed higher SFR at redshifts $z\gsim1-2$, compared to local rates. Nevertheless, this is the first time it has been measured, supporting the general shape of the star formation history as predicted from e.g., measurements of galaxy UV-luminosities. We revisit this discussion below.

\section{Relation between Star Formation and Supernova Rates}
With sufficient knowledge of how a star becomes a SN, it is possible to calculate the relation between the SFR and the SNR. Depending on the nature of the relation, it is possible to either use the SNR as a probe for the SFR, or to use independently determined SFRs and SNRs to investigate the relation between the two and extract information about the nature of the SN progenitors. For CC SNe, the relation between SFR and SNR is fairly straightforward since the progenitors are massive stars with life times spanning only a small fraction of the Hubble time. Therefore, if the mass range of CC SN progenitors and the IMF are known, we can calculate a constant that relates the SFR to the SNR. For type Ia SNe, the case is more complicated since there is a more substantial delay time involved, representing the time elapsed between the formation of the progenitor star and the explosion of the SN Ia. The distribution of delay times is not well understood, but is expected to be a large enough fraction of the Hubble time to untie any direct relation between SFR and SNR. Also, the efficiency, i.e., fraction of available progenitor white dwarfs that actually explode as SN Ia, is unknown.

In this paper, we use the recently determined SFR in Giavalisco et al. (2004b) as our fiducial model (M1) when investigating the relation between observed SNR and underlying SFR. In Giavalisco et al., the SFR at $z>3.5$~is determined using GOODS/ACS data, while rates from the literature are used at lower redshift (Lilly et al. 1996; Connolly et al. 1997; Steidel et al. 1999). The SFR, which is derived from UV-luminosity densities, after correcting for dust extinction, is characterized by a factor $\sim 10$~increase between $z=0$~and $z\sim 2$, thereafter the rate declines slowly toward higher redshifts up to $z\sim 6$. The SFR fitted to these data is shown in Figure 8 in Strolger et al. (2004). For comparison, we also fit data points in Giavalisco et al. without corrections for dust extinction. This model (M2) shows a similar increase in the SFR to $z\sim 1.5$ as in model M1. At higher redshifts, model M2 shows a decrease in the SFR, similar to the shape of the SFR first suggested by Madau et al. (1996).

\subsection{Star Formation Rates from Supernova Rates}
The progenitors of CC SNe are believed to be massive stars in the range $8M_{\odot}\lsim M\lsim 50M_{\odot}$ (e.g., Nomoto 1984; Tsujimoto et al. 1997). For an assumed initial mass function (IMF), $\psi(M)$, the number of CC progenitors per unit formed stellar mass is
\begin{equation}
k=\frac{\int^{50M_{\odot}}_{8M_{\odot}}\psi(M)dM}{\int^{125M_{\odot}}_{0.1M_{\odot}}M\psi(M)dM}
\end{equation}

Evaluating equation (2), assuming a Salpeter IMF (Salpeter 1955), yields $k$=0.0069 $M_{\odot}^{-1}$. Since the life time of the progenitor star is sufficiently short, we can assume a direct
relation between the mass of formed stars (SFR in units $M_{\odot}$yr$^{-1}$Mpc$^{-3}$),
and the number of exploding CC SNe (SNR in units yr$^{-1}$Mpc$^{-3}$),
\begin{equation}
SNR(z)=k\times h^2\times SFR(z)
\end{equation}
In this paper, we compare our estimated SNRs with SFRs derived from observed luminosity densities. Calculating the SFR from luminosity densities results in a rate proportional to the Hubble constant to the power one, i.e. $\propto h$. However, measuring SNRs from counting SNe in a volume, results in a rate that is proportional to $h^3$. Therefore, rates measured in these two ways will differ by a factor $h^2$, as shown in equation (3).

\begin{figure}
\epsscale{0.6}
\plotone{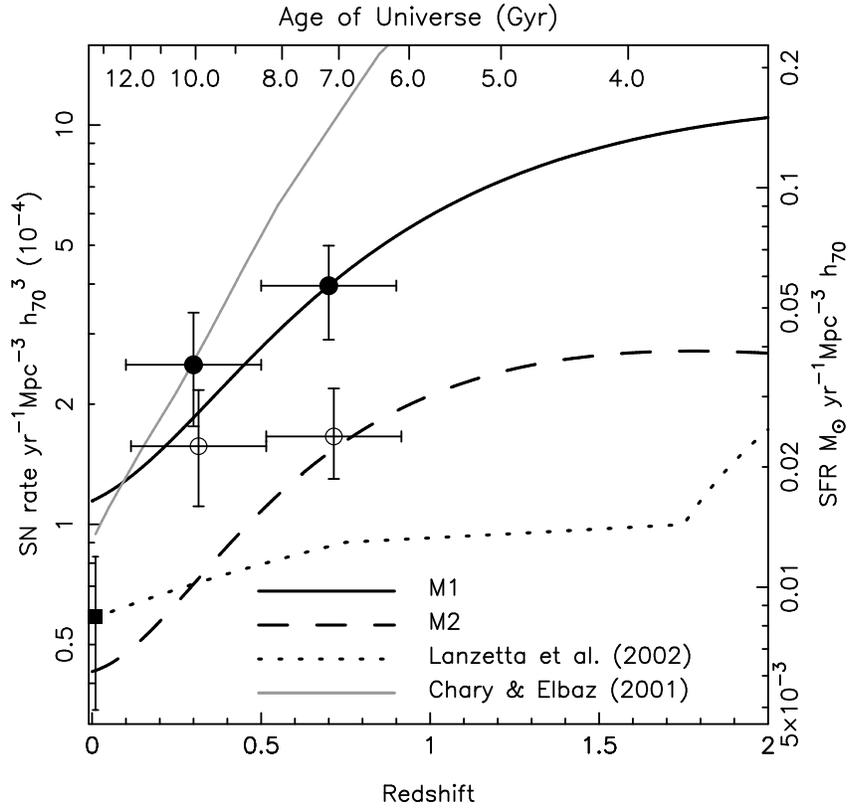}
\figcaption[f1.eps]{Core collapse supernova rates from GOODS are shown as filled circles. Vertical error bars show statistical errors, while horizontal error bars show bin size. Star formation model M1 (solid black line) is derived from the data in Giavalisco et al. (2004b), and uses similar assumptions about dust extinction as applied here. The star formation rate derived from analysis of various mid- and far- infrared datasets by Chary \& Elbaz (2001) is shown with the solid gray line. Also shown are star formation model M2 (dashed line) which is also derived from the Giavalisco et al. data and a star formation rate model given in Lanzetta et al. (2002) (dotted line). The latter two models are not corrected for dust extinction. For a comparison between these and our results, we show as open circles the rates derived assuming no dust extinction. The local rate from Cappellaro et al. (1999) is shown as a filled square.
\label{figure1}}
\end{figure}

In Figure \ref{figure1}, we plot our measured rates at $<z>=0.3$~and $<z>=0.7$ as filled circles, together with the local rate at $z\sim 0.0$ from Cappellaro et al. (1999) (filled square). The SNR in Figure \ref{figure1} is given by the scale on the left y-axis, while the corresponding SFR is given by the right y-axis. The relative off-set between the scales is given by the factor $k\times h^2$~in equation (3). Also plotted are the two different SFR models described above. The SFR derived from the CC SNe is slightly higher, but still consistent with the SFR derived from extinction corrected UV-luminosities shown as model M1. Based on our two data points, we find that the SFR increases by a factor $\sim$~1.6 between $<z>=0.3$~and $<z>=0.7$. However, a flat rate over this range is consistent within error-bars. Including the local rate from Cappellaro et al. (1999), we find that the SN based SFR shows a steep (factor $\sim$~7) increase in the redshift range $0<z\lsim0.7$, similar to model M1. In more detail, a chi-square fit shows that the SFR derived from SNe is on average $\sim$~13\% higher than the SFR in model M1, this difference is, however, within errors and may not be significant. Shown in the figure is also the dust enshrouded SFR derived from analysis of various mid- and far- infrared datasets (Chary \& Elbaz 2001). We hereafter refer to this as the IR SFR. The IR SFR matches the SN derived SFR at $z\sim$~0.3, but is a factor $\sim$~2 higher at $z\sim$~0.7. Considering errors, we conclude that both the IR and the extinction corrected UV-luminosity derived SFRs are consistent with our data points. We also note that if we combine the local rate from Cappellaro et al. (1999) with our data points, we can reject a flat CC SNR with 99.9\% probability ($\chi^2/\nu=7.3$), considering statistical errors only. The conclusion that the rate increase with redshift, therefore seems robust. We further investigate this conclusion in \S 5.1.1, where systematic errors discussed. 

For comparison, we also plot the SNR assuming no dust extinction (Figure \ref{figure1}, open circles, see also Table 2). This is certainly not physically correct since we know that the light from CC SNe is at least partly extinguished, but it shows an absolute lower limit to the rates. Despite the absence of correction, the SNR we measure is still higher than the corrected local rate. The SFR derived from the uncorrected SNR approximately follows the shape of the uncorrected SFR model M2, but is on average a factor $\sim$1.5 higher. Finally, we also plot in Figure \ref{figure1} the SFR derived by Lanzetta et al. (2002). We have here taken their 'middle' model and transformed rates to the cosmology used here. No correction for extinction is made in Lanzetta et al., we therefore compare their rates with the SFR derived from the uncorrected SNR. We find that the rates in Lanzetta et al. are a factor $\sim 2$ lower than the rates we find, this difference is at a $\sim 3\sigma$~level (99.6\%) considering statistical errors only. There are alternative models in Lanzetta et al. with somewhat higher SFR, however, all models predict rates lower than found here.

\subsubsection{Systematic Effects in CC SN Rates}
When calculating the CC and type Ia SN rates our results depend on a correct type determination of our SN sample. As mentioned in \S~2.2, we are most confident in the type determination of 7 out of 17 of the CC SNe and 20 out of 25 of the Ia SNe (called 'Gold' and 'Silver' SNe in Strolger et al. 2004). To investigate possible systematic effects that misclassification may introduce, we calculate errors in the rates assuming that all SNe that do not have absolutely certain determination of type have been assigned a wrong type (i.e., 'Bronze' type SNe in Strolger et al.). The systematic errors due to possible misclassification are 2.51$~^{+0.27}_{-1.38}$~and 3.96$~^{+0.93}_{-1.61}$ in the two bins, respectively (units are $10^{-4}$yr$^{-1}$Mpc$^{-3}h_{70}^3$). Inspecting the errors shows that statistical errors most often dominate over these systematic errors. There are, however, cases where the latter dominate. Note that the systematic errors due to possible misclassification are not 1$\sigma$-errors, but represent a worst case scenario, i.e., where all SNe that we are not certain of, are misclassified. In reality, we expect that at most a few SNe are misclassified.

Furthermore, when estimating the SNR, as well as a SFR from the SNR, our calculations include a number of assumptions and SN properties that, to some extent, are poorly known and may introduce additional systematic errors. First, the SNR depends on a number of characteristics, including light curves, peak magnitudes, K-corrections and dust extinction. Second, when deriving the SFR from the SNR, we have to specify the constant $k$, as well as the Hubble constant, in equation (3).

To determine how sensitive our results are to the assumptions that go into the calculations, we investigate how much we have to vary the assumptions in order for the results to change by more than 1$\sigma$ of the statistical errors ($\sigma_{stat}$). For CC SNe, $\pm 1\sigma_{stat}$~corresponds to $\pm$30\% in the estimated rates (using approximately the mean of the statistical errors of the two data points). After this investigation, we make an estimate of the total systematic error. We first investigate possible sources for systematic errors in the derived CC SNR.
\begin{itemize}
\item {\it Peak magnitudes}.
CC SN peak magnitude distributions are subject to some uncertainty because deriving these from observations depend on assumptions about extinction (as well as observational uncertainties). The typical uncertainty in peak magnitudes for different CC subtypes in Richardson et al. (2002) is 0.1--0.2 mag. In order to change the estimated number of CC SNe with $\pm\sigma_{stat}$, we have to change the assumed peak magnitudes of all subtypes, and in the same direction, by +0.3/--0.6 mag, which is unlikely considering the typical peak magnitude errors quoted. To further investigate this, we rerun our MC simulations incorporating the peak magnitude uncertainties for each subtype given in Richardson et al. We find that this only introduces an extra dispersion in the resulting rates by 2 and 4\% in the low and high redshift bins, respectively. This is insignificant compared to statistical errors.
\item {\it Subtypes}.
Changing the division into different subtypes may affect results. It is, however, not simple to quantify this in a direct way. First, we examine how the results change under the assumption that all CC SNe are of a single type represented by the SN IIP characteristics. We find that the resulting number of expected SN decreases by $\sim$15\%, which is less than the statistical errors and suggests that results are not highly dependent on the exact division into subtypes. The reason for this weak dependence is that omitting both the bright and the faint sub-populations cancels most of the net effect. What could change the results systematically, is if the fractions of very faint or very bright subtypes are significantly wrong. To examine the effects of overestimating either of these populations we calculate the predicted number of SNe after setting the fractions of bright and faint subtypes to zero, respectively. Excluding the bright Ibc and IIL population only changes the result by 3\%, while an exclusion of the faint IIP population changes the rate by at most 15\%, both changes are less than statistical errors. Next we examine how much we have to increase the fractions of faint or bright SNe in order to change the predicted number by $\sigma_{stat}$. We find that the intrinsic fraction of bright SNe have to increase from 1\% to $\sim$25\%, while the faint fraction have to increase from 15\% to $\sim$35\%. Neither of these scenarios seem likely.  
\item {\it Dust extinction}.
The resulting rates are dependent on the assumptions about dust extinction, as shown in Figure \ref{figure1} where our extinction corrected rates are a factor $\sim$~2 higher than the uncorrected. In order to change our results by $\pm\sigma_{stat}$, we have to change the assumed mean $E(B-V)$=0.15~with $\pm \Delta E(B-V)$=0.06. Using a Galactic extinction law ($R_V=3.1$) instead of a Starburst ($R_V=4.05$) decreases the estimated rates by $\sim$~9--15\%, significantly less than statistical errors.
\end{itemize}
\noindent
Besides these possible systematic effects when deriving the SFR, there are also systematic effects that may enter the derivation of the SFR from the SNR due to uncertainties in $k$~and $h$~in equation (3). 
\begin{itemize}
\item {\it IMF}.
The constant $k$~in equation (2) depends on the IMF and the assumed mass range for the progenitor stars. Changing the shape of the IMF assuming a Scalo IMF (Scalo 1986) instead of the Salpeter IMF, decreases $k$ by a factor 2.6. However, changing the IMF also changes the conversion from observed UV-luminosities to estimated SFRs. For a Scalo IMF, the conversion factor increases by a factor $\sim$2, i.e., the net effect of changing the IMF is therefore a decrease in $k$ by $\sim$30\%, which increases the measured SFR by the same amount. Again this does not exceed the statistical error in the derived rates.
\item {\it Progenitor mass range}.
We have chosen 50$M_{\odot}$ as the upper limit to the progenitor mass since it is believed that more massive star become black holes without exploding as SNe (Tsujimoto et al. 1997). Due to the steep slope of the IMF, increasing the upper limit to 125$M_{\odot}$, only increases the factor $k$~by 7\%. If we instead lower the upper limit to 30$M_{\odot}$, we get a resulting decrease in $k$ by 9\%. A more significant change in $k$ comes from changing the lower mass limit of the progenitor stars. To change $k$~corresponding to $\sigma_{stat}$, we have to increase the lower integration limit to 10.2$M_{\odot}$. This could be possible since the lower mass range of CC SNe is not well constrained and is estimated to be in the interval $8-11M_{\odot}$~(Timmes, Woosley, \& Weaver 1996). 
\item {\it Hubble constant}.
Finally, the relation between SNR and SFR depends on the Hubble constant to the second power. According to Spergel et al. (2003), the error in the determination of $H_0$~is less than 5\%, suggesting that the uncertainty due to the $h^2$~factor in equation (3), should be $\sim$10\%, clearly less than statistical errors. Note that the comparison between the SFRs derived from luminosity densities and SNRs respectively, provides a method for determining $H_0$. For a stringent determination of $H_0$, both statistical and systematic errors must, however, be significantly better constrained than is possible today. At face values, we can take our observed rates and compare to the SFR derived from UV-luminosity densities to calculate $H_0$. This exercise results in $H_0$=66$\pm 8$~km s$^{-1}$~Mpc$^{-1}$, where errors are statistical and represent 1$\sigma$.      
\end{itemize}

Errors introduced by systematic effects described above are mostly non-Gaussian, and it is difficult to estimate the total effect of the uncertainty on the results. For the SNR, only assumptions about dust extinction may have a significant effect on the rates. When deriving the SFR from the SNR, we add sources with possible systematic effects. 
To estimate the total error, we make the assumption that the uncertainty in peak magnitudes introduces a 5\% error, while division into subtypes introduce an additional 15\% uncertainty. Furthermore, we assume uncertainties due to dust extinction 10\% and 20\% in the two redshift bins, respectively. This results in a total error in the SNR of $\sim 19-25\%$, equivalent to $\sim 0.6-0.8\sigma_{stat}$. For the SFR derived from the CC SNR, we add three possible sources of statistical errors as shown by the listed points above. Assuming that each of these introduce a 10-15\% uncertainty would lead to a systematic error that is of the same order as the statistical error ($\sim 1\sigma_{stat}$).

To investigate if the strong significance of the increased SNR in the redshift range $0<z<0.7$ found in \S 5.1 decreases when taking systematic effects into account, we assume systematic errors $\sim0.6-0.8\sigma_{stat}$, as derived above. In section \S 5.1, we found that a flat CC rate could be rejected with 99.9\% probability, based on statistical errors only. Adding the estimated systematic errors decreases this probability to 99.4\%, which is still significant. 

We also note that the systematic effects discussed above are mostly independent of redshift. This implies that the relative increase in SNR over the redshift range observed should remain the same, and therefore supporting an increase in SFR, even if the absolute normalization of our determined rates may have a systematic offset. Only the assumptions about the amount of dust extinction have a strong redshift dependent effect on the derived rates. Assuming a higher extinction results in a steeper increase with redshift of the estimated rates, as can be seen in Figure \ref{figure1} by comparing the corrected and uncorrected data points.    

In summary, there are a number of possible sources for systematic errors in our estimates, however, most errors are relatively small, not exceeding the statistical errors. For the SNR, we estimate that the summed systematic errors should be smaller than the statistical, while the systematic errors may be of the same order as the statistical when it comes to the SFR derived from the SNR. Since we have shown that systematic errors are unlikely to dominate over the statistical, and that they are mostly independent of redshift, we are confident that the increase in supernova and star formation rates we observe are true features. The main concern is the amount of dust extinction. The effects of changes in the dust extinction are further discussed in \S 6. When we derive the SFR from the SNR, we also note that the direction of the errors is mainly to increase the observed rates, e.g., this is the case if the lower limit for CC progenitor mass is larger than 8$M_{\odot}$, if the IMF is changed to become less top heavy or if the amount of dust extinction is underestimated.

We finally note that the quoted systematic errors in Table 2 are sums
of the 19-25\% uncertainty derived above, and the uncertainty due to
possible misclassification. These errors are therefore non-Gaussian.

\subsection{Type Ia supernova rates}
Even though the physics behind type Ia SNe has been extensively investigated using both observations and theoretical simulations, there is still a lack of understanding of the mechanisms that proceed the explosion of this SN type (see Livio 2000, 2001 for reviews). The evolution of the type Ia SNR should follow the SFR, but shifted towards lower redshift after taking a delay time into account. For a particular distribution of delay times, $\Phi(t_d)$, where $t_d$ is the time elapsed between the formation of the progenitor star and the explosion of the SN Ia, the SNR is given by a convolution of the SFR over delay times
\begin{equation}
SNR_{Ia}(t)=\nu \int_{t_F}^t SFR(t^{\prime})\times \Phi(t-t^{\prime}) dt^{\prime}
\end{equation}
where $t$~is the age of the Universe and $t_F$~is the time corresponding to the redshift, $z_F$, where the first stars formed. In this paper we set $z_F$=10. $\nu$ is the number of SNe per unit formed stellar mass ($M_{\odot}^{-1}$). In Strolger et al. (2004), we calculate predicted redshift distribution assuming two different functional forms for the distribution of delay times. The first model has an $e$-folding delay time distribution, $\Phi(t_d)\propto exp(-t_d/\tau)$, while the second model has a Gaussian distribution, $\Phi(t_d)\propto exp\{-[(t_d-\tau)/2\sigma]^2\}$. We call $\tau$~the characteristic delay time. Note that this parameter has different meanings in the two functions. The $e$-folding distribution always has the highest probability for $t_d$=0, and has $\sim 1/3$~of the SNe with $t_d > \tau$. The Gaussian distribution has the highest probability for $t_d = \tau$, and $1/2$~of the SNe have $t_d > \tau$. We use two different Gaussian models, one 'narrow' with $\sigma = 0.2\tau$~and one 'wide' with $\sigma = 0.5\tau$. By comparing the observed redshift distribution with the predicted, we investigate in Strolger et al. (2004) which shapes of the delay times function, as well as which ranges of $\tau$~are consistent with the observed distribution. We find that all models favor characteristic time scales, $\tau$, around or greater than $\sim$ 3 Gyr. Models with delay times $\tau < 2$~Gyr can be rejected with 95\% confidence. These constraints are similar to the ones found by Gal-Yam \& Maoz (2004), who estimate the delay time using an $e$-folding distribution only. Using a different approach, Maoz \& Gal-Yam (2004) argue that the delay-time must be short if the iron mass observed in clusters of galaxies originate from SN Ia. Further comparisons between these investigations are presented in Strolger at el. (2004).

\begin{figure}
\epsscale{0.6}
\plotone{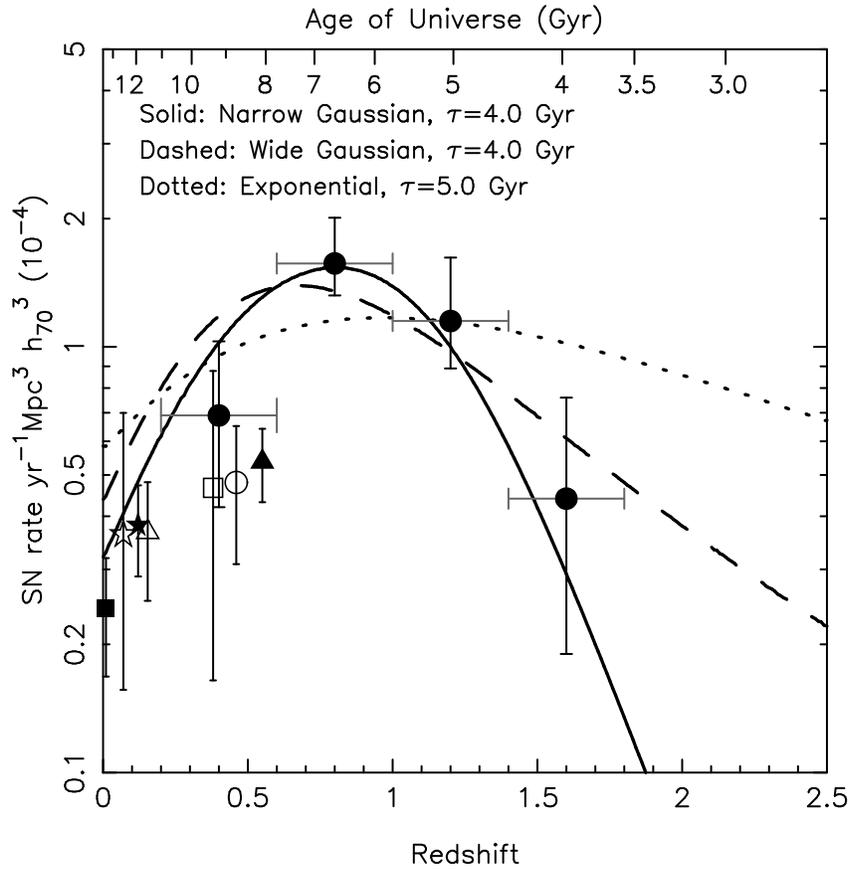}
\figcaption[f2.eps]{Type Ia SNRs from GOODS at $<z>$=0.40, 0.80, 1.20, and 1.60 are shown as filled circles. Also plotted are rates at $z\sim$~0.01 from Cappellaro et al. (1999)(filled square), at $z\sim$~0.1 from Hardin et al. (2000)(open star), at $z\sim$~0.11 from Strolger et al. (2003a)(filled star), at $z\sim$~0.11 from Reiss (2000)(open triangle), at $z\sim$~0.38 from Pain et al. (1996)(open square), at $z\sim$~0.46 from Tonry et al. (2003)(open circle), and at $z\sim$~0.55 from Pain et al. (2002)(filled triangle). Vertical error-bars on the GOODS rates represent statistical errors, while horizontal error-bars represent bin size. See text for discussion on systematic errors. The figure also shows predicted Ia rates based on three different models for the delay time distribution of SN Ia progenitors.
\label{figure2}}
\end{figure}

In Figure \ref{figure2}, we plot observed data points from this investigation together with rates from Cappellaro et al. (1999), Hardin et al. (2000), Strolger (2003), Reiss 2000, Pain et al. (1996, 2002) and Tonry et al. (2003). We have converted the first four rates from SNu's using a $B$-band luminosity density given in \S~3. We also show SNRs derived from equation (4) using the three different delay time models. For the 'narrow' and 'wide' Gaussians, we use the best-fitting value $\tau=4.0$~Gyr, derived from fitting the models to the GOODS dataset in Strolger et al. (2004). For the $e$-folding we choose $\tau=5.0$~Gyr (there is no best-fitting value for this functional form, but values below $\sim$2 Gyr are rejected with 95\% confidence). We further use SFR model M1 in equation (4). In Strolger at al., we find that the 'narrow' Gaussian delay time distribution fits the GOODS data better than the other two models. From Figure \ref{figure2}, we see that this fit is also consistent with the low redshift ($z\lsim 0.1$) data taken from the literature. Only the two data points at $z\sim$0.5 deviate significantly from the fit and are about a factor $\sim$~2 lower than the model suggests. A possible explanation for this deviation is that ground-based searches may suffer from unaccounted incompleteness. B. Barris (2003, private communication), through a reanalysis of light curves of all variable objects, has found significant incompleteness in the ground based 2001 survey for SN Ia reported by Barris et al. (2004). As many as half of the bona fide SN Ia at $z<0.6$ were not recognized as such during the campaign. The salient characteristic of many of these missed SN Ia is extremely close proximity to a relatively bright galaxy which causes them to be identified as potential AGN, making them difficult to observe spectroscopically regardless of their true nature. It is not possible to know how much of this bias was present in the preceding Fall 1999 campaign reported in Tonry et al. (2003), but it indicates that the rate estimates contained therein could be low by up to nearly a factor of two, which would bring the rate at $z=0.5$ into much better agreement with that found here.

The reduced chi-squares when fitting the GOODS data points to models are $\chi^2/\nu$=0.46, 1.01 and 1.46, for the 'narrow' Gaussian, 'wide' Gaussian and e-folding distributions, respectively. The 'narrow' Gaussian has the lowest value since it best reproduces both the sharp increase at $z<1$ and the decline at $z>1$. The reduced chi-squares for the same distributions, including also rates from literature are $\chi^2/\nu$=7.4, 7.9 and 10.6, respectively. The relatively large chi-squares are mostly due to the deviant points at $z\sim$~0.5. Excluding these points results in $\chi^2/\nu$=1.1, 3.2, and 6.3, respectively. A similar improvement is also achieved if the $z\sim$~0.5 points are multiplied by a factor $\sim$~2. These results again favor the 'narrow' Gaussian as the best model. In Strolger et al., we concluded that delay time models predicting a large fraction of SN Ia with delay times shorter than $\sim$2 Gyr are inconsistent with data. Any such model does not follow the decline at $z>1$~that our observations suggest. 

\subsubsection{Systematic effects in Ia rates}
The measured type Ia rate is basically affected by the same systematic effects as the CC rate. First we have possible errors due to misclassification of SN type. These errors alone are 0.69~$^{+1.43}_{-0.14}$, 1.57~$^{+0.47}_{-0.26}$, 1.15~$^{+0.08}_{-0.20}$~and 0.44~$^{+0.04}_{-0.01}$, in the four redshift bins, respectively (in units $10^{-4}$yr$^{-1}$Mpc$^{-3}h_{70}^3$). For the remaining sources of errors, there are a few factors that make the SN Ia rates less subject to systematics compared to CC SNe. Characteristics like peak magnitudes, light curve shape and spectra are more universal, as well as better known, for Ia SNe, and should not introduce any significant systematics. The extinction corrections could introduce systematics that significantly affect the derived rates. However, the dependence on assumed extinction is likely to be smaller than for CC SNe since the type Ia SNe on average occur in environments that are less affected by dust extinction. To examine the sensitivity to systematic effects, we examine the type Ia SNe estimates in a similar way to the CC SNe. For the type Ia SNe we investigate how much we have to change the assumptions that go into our calculation in order to change the estimated rates by $\pm 1\sigma_{stat}$.  
\begin{itemize}
\item {\it Peak magnitude}.
To change the estimated rates in each of the four redshift bins with +1$\sigma_{stat}$, the peak magnitudes have to be more than 0.5 mag fainter than the values given in Table 3. Alternatively, the peak magnitude have to be more than one magnitude brighter to change the estimated number by $-1\sigma_{stat}$ in each bin. These required shifts are significantly larger than the accuracy with which the (mean) peak magnitude of Ia SNe is known.
In this investigation, we do not include the observed effect that brighter SNe Ia have a slower evolution on the light curve, while fainter SNe evolves faster. To a first approximation, these two effects cancel out, leaving no net effect on the derived rates. However, to investigate if there is a net effect at high redshifts where only the brightest SNe are expected to be detected, we run a set of MC simulations accounting for this. We use the relation between peak magnitude and light curve stretch given in Perlmutter et al. (1997). Repeating the simulations above, we find than the derived rates in the four redshift bins increase by less than 4\%, with the smallest increase in the highest redshift bin (1\%). Therefore, we find that the peak magnitude light curve stretch relation has little effect on the derived rates.
\item {\it Subtypes}.
Setting the fraction of either the faint or bright SN Ia subtypes to zero does not change the estimated rates by more than $\sim$~10\%. To change the estimated rates by 1$\sigma_{stat}$, we have to increase the fraction of faint SN Ia from 16\% to $\sim$~50\%. Setting the fraction to 100\%  for the bright population changes the rates by less than 1$\sigma_{stat}$. These calculations show that the division into subtypes should not introduce systematics that are comparable to statistical errors.
\item {\it K-corrections}.
We use the spectrum of SN1994D to calculate K-corrections since this is the only observed SN with sufficient spectral coverage available over a large number of epochs on the light curve. The good UV-coverage of SN1994D is essential for deriving the K-correction to redshifts $z\sim 1.8$. SN1994 is, however, about $\sim$0.3 mag bluer in $(U-B)$~color than the mean $(U-B)$~color in Leibundgut et al. (1988). To investigate if the unusually blue color of SN1994D affects results, we rerun our MC simulations with a magnitude correction to the K-corrections. We start to add the correction at the redshift where the effective wave-length of the observed filter ($z$-band) reaches the effective wave-length of redshifted rest-frame $B$-band (i.e., $z\sim$ 1). The correction increases linearly to reach 0.3 mag at the redshift where the observed band probes the effective wave-length of the redshifted $U$-band (i.e., $z\sim$ 1.6). At higher redshift, the correction is set to a constant 0.3 mag. We find that the effect of this correction on results is small. The derived rates increase by less than 5\% in the two highest redshift bins when including this color correction (bins at $z<1$~are not affected at all). We therefore conclude that the use of the unusually blue SN1994D when deriving K-correction should not affect results more than marginally. 
\item {\it Dust extinction}.
Using only half the nominal amount of dust extinction reduces the estimated rates in the four redshift bins by 5, 10, 14 and 19\%, respectively. All these changes are less than statistical errors. To increase the estimated rates by 1$\sigma_{stat}$, we need to increase the dust extinction by a factor $\sim$~3, compared to the extinction model in Hatano et al. (1998), used here. 
\end{itemize}
Based on the discussion above, we find that the systematic errors in the type Ia rates should not exceed statistical errors. Making the simple assumption that the three first listed sources introduces uncertainties $\sim$10\%, $\sim$5\%, and $\sim$5\%, while the uncertainty due to dust extinction increases from 10\% in the lowest redshift bin to 20\% in the highest redshift bin, results in a total added error $\sim16-23$\%, which is less than statistical errors in all bins, and is on average less that 0.6$\sigma_{stat}$. We are therefore confident that general evolution in the Ia rate we observe is a true feature. Quoted systematic errors in Table 2 include
these uncertainties as well as uncertainties due to possible misclassification of SN type.   
\subsection{White dwarf explosion efficiency}
Fitting the predicted model distributions of SN Ia to the observed sample includes determining the normalization, $\nu$, in equation (4). This number tells us how many type Ia SNe explode per unit formed stellar mass. Previously in this investigation, we have made no assumptions about the mass range of type Ia progenitor stars. If we make an assumption constraining the progenitor mass range, we can calculate the fraction of stars in this mass range that subsequently explode as SN Ia, which we here call the efficiency $\eta$, via the relationship
\begin{equation}
\nu=\eta \frac{\int^{8M_{\odot}}_{3M_{\odot}}\psi(M)dM}{\int^{125M_{\odot}}_{0.1M_{\odot}}M\psi(M)dM}
\end{equation}
The mass range of progenitor stars is set to $3M_{\odot}<M<8M_{\odot}$~(Nomoto et al. 1994). By fitting each of the three delay time models, $\Phi (t_d)$, in equation (4), we estimate $\nu$=1.0$\times 10^{-3}$, 1.2$\times 10^{-3}$, and 1.3$\times 10^{-3}$ for the 'narrow' Gaussian, the 'wide' Gaussian and the $e$-folding distributions, respectively. Evaluating equation (5) results in an efficiency of $\eta$= 4.9\%, 5.6\%, and 6.3\%, with which progenitor white dwarfs explode as SN Ia, depending on the delay time model. Furthermore, if we assume that all stars with main sequence masses $0.8M_{\odot}<M<8M_{\odot}$~become white dwarfs, then our observations suggests that $0.6-0.8$\% of the total number of white dwarfs will explode as SNe Ia. These numbers are consistent (within observational errors and the uncertainties associated with the progenitor models) with population synthesis calculations (e.g., Yungelson \& Livio 2000, and Yungelson, private communication), which predict this fraction to be $\sim$0.5\%.

\subsection{Ratio of the Core Collapse to Type Ia SN Rates}
\begin{figure}
\epsscale{0.6}
\plotone{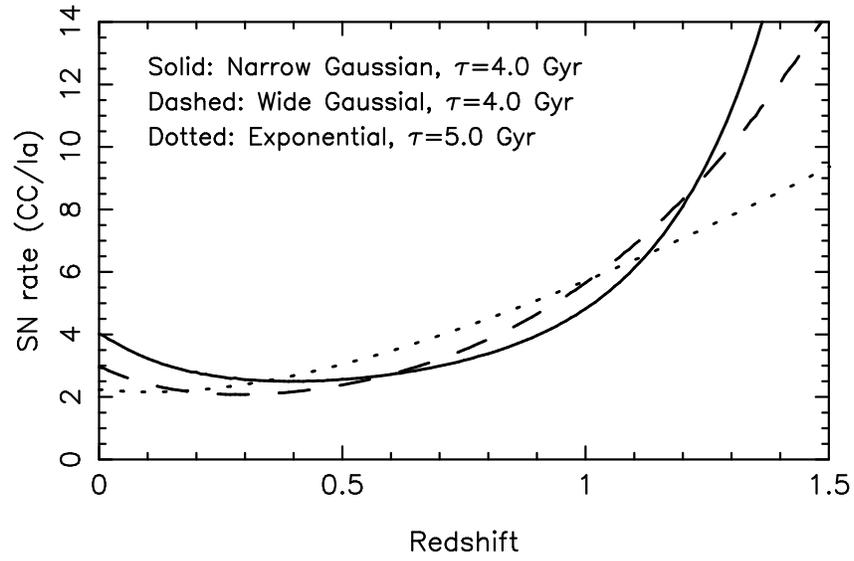}
\figcaption[f3.eps]{Intrinsic ratio of exploding core collapse to Ia SNe as a function of redshift. Ratio is shown for SFR model M1 using three different delay time distributions; the $e$-folding, the 'narrow' and the 'wide' Gaussian. The characteristic delay time for each model is shown in the figure caption.
\label{figure3}}
\end{figure}

In Figure \ref{figure3}, we plot the redshift evolution of the intrinsic ratio between CC and SN Ia rates, $r_{(CC/Ia)}$. For CC, we use rates calculated by fitting SFR model M1 to our observed data points. SN Ia rates based on the three delay time models, each having a characteristic delay time $\tau$ quoted in \S~5.2. For the 'narrow' and 'wide' Gaussian, we get local ratios between CC and Ia rates $r_{(CC/Ia)}\sim$~4.0 and 2.9, respectively. This is consistent with the local ratio $\sim$3.5 derived by Madau et al. (1998) using a compilation of locally determined rates. For the $e$-folding delay time models we get a local ratio 2.2. This lower value is mostly an effect of the higher type Ia rates at low redshift predicted in this model. The CC to Ia ratio stays fairly constant at a value $\sim$~3 to $z\sim$~0.7. The ratio increases rapidly at higher redshifts. The reason for this behavior is that the relatively long delay time shifts the SN Ia to lower redshifts compared to the SFR (and hence CC SNR). Therefore, SN Ia becomes relatively more common at lower redshift, lowering the CC to Ia ratio. In the case of a short delay time, the Ia SNR and the SFR should closely follow each other, giving a ratio $r_{(CC/Ia)}$~that would be relatively constant with redshift. Using MC simulations, we find that the statistical uncertainty in the derived ratio is $\sim\pm$30\%. Systematic errors add to these, however, the systematic errors are mostly independent of redshift and should therefore only change the normalization of the ratio and not the evolution with redshift. Therefore, the result that the ratio is fairly constant to $z\sim0.7$ and thereafter increases, should be robust. Note that these results refer to intrinsic ratios of exploding SNe. In a magnitude limited search, the detected ratio will be lower since CC SNe are typically fainter and are on average more severely affected by dust extinction.

\section{Discussion}
We have derived a SFR to $z\sim$~0.7 from measurements of the CC SNR that is $\sim$~13\% higher than (but consistent with) the SFR based on evolution of extinction corrected UV-luminosity densities. There are claims that a large fraction of the cosmic SFR is hidden from measurements based on rest-frame UV-luminosities. Driver (1999) argues that extreme selection effects bias against finding low surface brightness galaxies at $z~>~0.2$. These systems may therefore contain a large amount of undetected star formation. SNe occurring in low surface brightness environments are readily detected. In fact, SNe in these environments should be easier to detect since there is no noise added due to a background source when the host galaxy is undetected. However, we do not find more than one CC SN (out of 17) that lacks an apparent host galaxy. This suggests that only a limited amount of star formation may be hidden in such environments. The lack of host-less SNe also suggests that there is not a large amount of star formation occurring in faint low mass systems at high redshift as is the case locally (Cowie et al. 1996). 

It is difficult to estimate corrections due to dust extinction in the host galaxies. Because SNe are detected in the optical, instead of rest-frame UV, they are less affected by dust extinction, however, corrections are still sufficiently large to change the estimated rates considerably, as shown in \S~5. Observations of the luminosity density at longer wavelengths are less affected by dust extinction and may therefore offer a way to more directly determine the star formation. Chary \& Elbaz (2001) derive the SFR from mid- and far-IR observations and find a rate that follows the overall {\em shape} of the UV-determined rate, with a sharp increase to $z\sim~1.5$, and thereafter a constant or declining rate, similar to our model M2. The rate derived from the IR is, however, in general significantly higher than UV derived rate, a difference that increases with redshift. The IR derived SFR is a factor $\sim$~2 higher at $z=0.5$~and a factor $\sim$~3 higher at $z=1$, as compared to the extinction corrected UV-derived SFR from Giavalisco et al. (2004b). Our results from CC SNe match the IR SFR at $z=0.3$, but is a factor $\sim$2 lower at $z=0.7$. The reduced chi-squares when fitting the UV as well as IR derived SFRs to the observed points are 0.5 and 16, respectively, implying that the UV-derived SFR fits the observations better. However, an increased amount of dust extinction in the host galaxy will make observations more consistent with the IR data. To investigate this, we adjust the dust extinction so that the SFR derived from CC SNe fits the IR SFR. We find that a mean extinction $E(B-V)$=0.40 (instead of the nominal $E(B-V)$=0.15), results in the best fit between our SN derived SFR and the IR SFR. The reduced chi-square using this extinction is 3.5. 

So far, we have assumed a constant $E(B-V)$~over the redshift range probed. It is, however, possible that $E(B-V)$~evolves so that extinction increases with redshift. This was suggested by Totani \& Kobayashi (1999) who derived an evolution in A$_{\rm V}$ of $\sim$~0.2 mag between $z\sim$~0 and $z\sim$~1 when integrated over all galaxy types. Fitting our two data points independently results in best-fitting values $E(B-V)$=0.16 at $z$=0.3 and $E(B-V)$=0.50 at $z$=0.7, which corresponds to an evolution in A$_{\rm V}$ of $\sim$~1.4 mag in this redshift range. This is higher than the evolution found by Totani \& Kobayashi, but is more consistent with the analysis of Chary (2004) who has re-calculated the average extinction as a function of redshift based on the most accurate estimates of the luminosity density at far-infrared and ultraviolet wavelengths and finds that the best-fit $E(B-V)$ evolves from 0.19 at $z$=0.3 to 0.3 at $z$=0.7.

Finally, a large fraction of star formation may be hidden in extremely extinguished environments such as cores of starbursts or ULIRGs (Blain et al. 1999). It is hard to use optically detected SNe to probe this activity, since the absorption in these systems is typically A$_{\rm V}~\sim$~10 (Mannucci et al. 2003). Therefore, SFRs based on both UV-luminosities and optically detected SNe, will miss most of this population. Targeted observations, searching for SNe in the NIR, may be the way to detect a possible abundant population of CC SNe in these objects (Mattila \& Meikle 2001). Note that for type Ia SNe it is possible that the delay-time makes detection possible if sufficient time is available for the dust to settle or disperse before the explosion of the SN.  

We have found good agreement between the SFRs derived from SNe and galaxy luminosity densities. We do, however, note that there may be corrections to the derived rates. With the systematic effects being of the same order of magnitude as the statistical errors, it is necessary to increase the statistical sample, as well as learn more about systematic effects, in order to significantly reduce errors. Note, however, that uncertainties involved in deriving the SFR from high redshift galaxy UV-luminosities are also most certainly affected by systematic effects and unknown details regarding e.g., dust extinction and IMF, leading to uncertainties that should be of the same order of magnitude. Therefore, measuring the SFR from observations of CC SNe, provides an important independent probe for the cosmic SFR. Furthermore, compared to measurements of the SFR based on UV-luminosities, the point source SNe are not affected by surface brightness dimming effects, and are also somewhat less dependent on extinction since observations mainly probe rest-frame optical instead of UV. It is therefore of high interest to extend SN searches to larger areas, or number of observed epochs, to increase the sample and hence minimize the statistical part of the errors. A caveat is, however, that CC SNe are relatively faint, and it will be challenging to derive a sufficiently large, as well as complete, sample at redshifts exceeding $z\sim~1-2$, until the next generation of telescopes are operational. 

The results from type Ia rates clearly show an increase in the rate to $z\sim$~1, thereafter the rate flattens and shows a suggested decrease in the redshift range $1\lsim z\lsim1.6$. Deriving a SNR with this shape from a SFR scenario that is consistent with recent observations of the UV-luminosity density, requires that there is a significant delay between the star formation and the subsequent explosion of the Ia SNe. In Strolger et al. (2004), we showed that a 'narrow' Gaussian distribution of delay times best fits our observations. However, other functional forms for the delay time may also be consistent with observations as long as the characteristic time delay, $\tau$, is $\gsim$~2 Gyr. Here we have shown that the 'narrow' Gaussian distribution is best-fitting also when we enlarge our sample of observed rates by rates taken from the literature.

To further constrain the characteristic time delay $\tau$, as well as the functional forms of the delay time, it is necessary to determine the shape of the SFR to greater detail than known today. Especially, an increased redshift coverage will make it possible to set firmer constraints on $\tau$. This is shown in Figure \ref{figure2}, where models diverge significantly at higher redshift $z\gsim$~2. Alternatively, if studies of SN Ia can reveal the mechanisms that lead to the explosion, i.e., determining the distribution of delay times, then it will be possible to set firmer constraints on the SFR using the observed SNR. 
\section{Conclusions and summary}
Using a unique sample of high redshift supernovae obtained with the ACS on-board HST, we have calculated the rates of core collapse SNe to $z\sim$~0.7 and type Ia SNe to $z\sim$~1.6. These redshifts are significantly higher than those where rates have previously been measured. We find a core collapse SN rate that is a factor $\sim$~1.6 higher at $<z>\sim$~0.7 than at $<z>\sim$~0.3. Compared to local rates, we find an increase in the rate by a factor $\sim$~7 over the redshift range $0<z<0.7$. Relating the SN rate with the star formation rate, we find that this increase is consistent with the evolution of the cosmic star formation rate measured from extinction corrected galaxy UV-luminosity densities. The rates we derive are also consistent with the star formation rate estimated from mid- and far-IR luminosity densities, even though the data do not fit as well as with the UV-luminosity density. A better fit with the IR derived star formation rate would be achieved in a scenario where the dust extinction increases with redshift.

These first measurements of the SFR based on CC SNe are affected by both statistical uncertainties and possible systematic errors, as extensively discussed above. However, most of these errors are independent of redshift. Therefore, the relative increase in the SFR with redshift is robust.

The type Ia SN rate shows a significant increase to $z\sim$~1, compared to measurements at $z<0.5$ taken from the literature. In the redshift range $1\lsim z\lsim 1.6$, we find a SNR that is consistent with decreasing. When fitting observed rates to those derived from observed star formation rates combined with different assumptions on the SN Ia delay time distributions, we find that the observed shape of the Ia rate is most consistent with models assuming a significant delay time between the epoch of star formation and the explosion of the Ia SN. We find characteristic delay times $\tau\gsim$~2 Gyr are consistent with data. This provides further support to arguments favoring single-degenerate scenarios for the progenitors of SNe Ia (Livio 2001).

Fitting observations to models also provide estimates of the fraction of formed stars that eventually explode as SN Ia. Assuming a Salpeter IMF and a progenitor main sequence mass range $3M_{\odot}<M<8M_{\odot}$, we find that $5-7$\% of the white dwarfs formed from these progenitor stars subsequently explode as SNe Ia. Furthermore, if we assume that all stars with main sequence masses $0.8M_{\odot}<M<8M_{\odot}$ become white dwarfs, then $0.6-0.8$\% of these will turn into SNe. Investigating the intrinsic ratio of exploding CC to Ia SNe results in a ratio $\sim$~3, which is fairly constant up to $z\sim$~0.7. At higher redshifts the ratio increases, reflecting the fact that a long delay time for Ia progenitors shifts the rate towards lower redshift compared to the SFR and therefore also the CC SNR. 

To set firm constraints on the SFR using core collapse SNe, as well as determining the distribution of delay times for SNe Ia progenitors, it is of great importance not only to increase the sample size (i.e. decrease statistical errors), but also to better understand SN characteristics. Nevertheless, using available information on SN statistics and characteristics, combined with the unique sample of high redshift SNe provided by the GOODS, we have already shown that using high redshift supernova rates can provide important information about the properties and the evolution of stars and star formation in the Universe at earlier cosmic epochs.

\acknowledgments{
We thank the GOODS team for excellent work during the supernova search campaign. We thank the referee Dan Maoz for valuable comments and suggestions for improving the manuscript. We also thank Brian Barris for discussion on completeness of ground based searches. Support for the GOODS HST Treasury program was provided by NASA through grants HST-GO-09425.01-A and HST-GO-09583.01 from the Space Telescope Science Institute, which is operated by the Association of Universities for Research in Astronomy, under NASA contract NAS5-26555. P.M. acknowledges support by NASA through grants NAG5-11513 and GO-09425.19A from the Space Telescope Science Institute.
}


\clearpage



\clearpage

\clearpage

\clearpage

\begin{deluxetable}{lccc}
\tabletypesize{\scriptsize}
\tablewidth{0pt}
\tablecaption{Supernova characteristics.}
\tablecolumns{4}
\tablehead{
\colhead{Type} & \colhead{$M_B$}& \colhead{$\sigma$}  & \colhead{$f$}
}
\startdata
\multicolumn{4}{c}{{\it Type Ia}}\\
Ia-'bright'        & -19.6 & 0.30 & 0.20 \\
Ia-'normal'        & -19.3 & 0.45 & 0.64 \\
Ia-'faint'         & -17.8 & 0.50 & 0.16 \\
\multicolumn{4}{c}{{\it Core collapse}}\\
Ibc                &  -17.28 & 0.74 & 0.225 \\
Ibc-'bright'       &  -19.93 & 0.33 & 0.005 \\
IIP                &  -16.66 & 1.12 & 0.300 \\
IIP-'faint'        &  -14.39 & 1.00 & 0.150 \\
IIL                &  -17.22 & 0.38 & 0.295 \\
IIL-'bright        &  -18.94 & 0.51 & 0.005 \\
IIn                &  -18.82 & 0.92 & 0.020 
\enddata
\tablecomments{Columns are: (1) Supernova type;
(2) Absolute peak $B$ band magnitude assuming $H_0$=70 km s$^{-1}$Mpc$^{-1}$ and a cosmology with $\Omega_M=0.3$~and $\Omega_{\Lambda}=0.7$;
(3) Dispersion in peak magnitude;
(4) Intrinsic fractions of subtypes within the two groups.
}
\end{deluxetable}

\clearpage
\begin{deluxetable}{lcccc}
\tabletypesize{\scriptsize}
\tablewidth{0pt}
\tablecaption{Supernova rates}
\tablecolumns{5}
\tablehead{
\colhead{} &\colhead{} &\colhead{} &\colhead{} &\colhead{}
}
\startdata
Redshift:  & $0.2\le z<0.6$ & $0.6\le z<1.0$  & $1.0\le z<1.4$ & $1.4\le z<1.8$ \\
{\it Type Ia} & 0.69~$^{+0.34~+1.54}_{-0.27~-0.25}$ & 1.57~$^{+0.44~+0.75}_{-0.25~-0.53}$ & 1.15~$^{+0.47~+0.32}_{-0.26~-0.44}$ & 0.44~$^{+0.32~+0.14}_{-0.25~-0.11}$ \\
{\it Number of SNe}& 3 & 14 & 6 & 2\\
 & & & & \\
Redshift:   & $0.1\le z<0.5$ & $0.5\le z<0.9$  & \\
{\it Core collapse}& 2.51$~^{+0.88~+0.75}_{-0.75~-1.86}$ & 3.96$~^{+1.03~+1.92}_{-1.06~-2.60}$  & & \\
\multicolumn{4}{c}{Rates without extinction correction} & \\
 & 1.57~$^{+0.60~+0.42}_{-0.46~-1.12}$ & 1.66~$^{+0.53~+0.65}_{-0.36~-0.93}$  & & \\
{\it Number of SNe}& 6 & 10 & &
\enddata
\tablecomments{Rates are given in units $yr^{-1}~Mpc^{-3}~10^{-4}~h_{70}^3$, assuming a cosmology with $\Omega_M=0.3$ and $\Omega_{\Lambda}=0.7$. First quoted errors are statistical and represent 68.3\% confidence intervals. Second errors are systematic and include the possibility that all SNe with uncertainty in type determination have been misclassified, as well as other possible sources discussed in \S 5.1.1 and \S 5.2.1. Note that the latter errors are non-Gaussian.
}
\end{deluxetable}


\end{document}